\documentclass[pre,superscriptaddress]{revtex4}
\usepackage{graphicx}
\usepackage{color}
\usepackage{verbatim}
\newcommand{\br}{{\bf r}}
\newcommand{\bk}{{\bf k}}

\newcommand{\bR}{{\bf R}}

\begin{document}
\title{Depletion interaction between spheres of unequal size and demixing in binary mixtures of colloids}
\author{Alberto Parola}
\affiliation{Dipartimento di Scienza e Alta Tecnologia,
Universit\`{a} dell'Insubria, Via Valleggio 11, 22100 Como, Italy}
\author{Luciano Reatto}
\affiliation{Via Bazzini 20, 20133 Milano, Italy
}
\begin{abstract}
The possibility to induce demixing in a colloidal mixture by adding 
small polymers, or other equivalent depletant agents, is theoretically investigated. 
By use of Mean Field Theory, suitably generalized to deal with short range 
effective interactions, the phase diagram of a binary mixture of 
colloidal particles (modelled as hard spheres) in a solvent is determined 
as a function of the polymer concentration on the basis of the Asakura-Oosawa model. 
The topology of the phase diagram changes when the relative size of the colloidal 
particles is reduced: the critical line connecting the liquid-vapour critical points of
the two pure fluids breaks and the branch starting from the critical point of the 
bigger particles bends to higher volume fractions, where concentration fluctuations
drive the transition. The effects of a softer colloid-polymer interaction is also investigated: 
Even the presence of a small repulsive tail in the potential gives rise to a
significant lowering of the stability threshold. In this case, phase transitions may take place 
by adding just a few percent of depletant in volume fraction.
These results may be relevant for the interpretation of recent experiments of 
solidification kinetics in colloidal mixtures.
\end{abstract}
\maketitle

\section{Introduction}
The realm of liquid state theory~\cite{hansen}, initially developed to study the so called 
simple fluids like those whose constituents are noble gases or small molecules, 
has been enormously extended when it was realized that such methods could be 
imported to treat many phenomena in a significant fraction of complex liquids, 
like polymers in solution or colloidal suspensions. The strategy is first to identify, 
among the plethora of the microscopic degrees of freedom, a few effective ones
and then to construct an effective interaction among these. Models like the hard sphere 
system or the Yukawa fluid found accurate experimental implementations~\cite{likos} 
in such context and the analog of the standard liquid-vapor and solidification phase 
transitions are quite common in such complex fluids. However, it should be kept 
in mind that the effective interaction can be very different from those typical 
of a simple fluid and the phase diagram can display more complex topologies compared to the textbook 
paradigm of three phases, vapor, liquid and solid, the last one occupying the high-density side. 
The liquid phase can disappear altogether or be present only as a metastable state~\cite{frenkel};
re-entrant solidification~\cite{likos2} may appear when a fluid region is present 
at higher density, and density modulated phases can also be found~\cite{seul}
in place of a liquid or of a solid.

\begin{figure}
\includegraphics[height=5cm,width=10cm,angle=0]{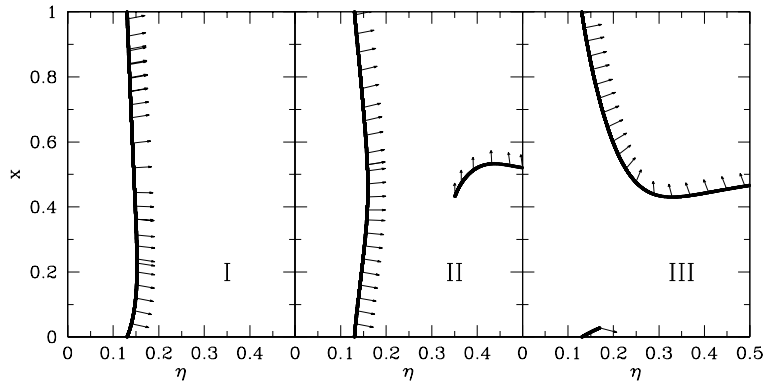}
\caption{Mean field critical lines projected
in the plane total packing fraction/volume fraction of component 1,
($\eta=\eta_1+\eta_2=\frac{\pi}{6}\rho_1+\frac{\pi}{6}\rho_2$, $x=\frac{\eta_1}{\eta}$)
for three Lennard-Jones (LJ) mixtures. The diameter ratios $\sigma_2/\sigma_1$ are 
$0.9$ for type I and type II systems and  $0.8$ for the type III model. 
In the three cases the ratios between the amplitudes of the LJ potentials are:
Type I, $\epsilon_{11}=1.7$, $\epsilon_{12}=1.3$;
Type II, $\epsilon_{11}=1.7$, $\epsilon_{12}=1.2$;
Type III, $\epsilon_{11}=3$, $\epsilon_{12}=1.7$ in units of $\epsilon_{22}$. 
The direction of the arrows identify the direction of the order parameter: an arrow parallel to the density axis
indicates a liquid-vapour transition, an arrow along the volume fraction axis, a demixing transition.
}
\label{Fig1}
\end{figure}

In the field of simple fluids when one moves from a one component system to a 
binary mixture a very rich phase diagram is found~\cite{rowlinson}  and some of them are sketched 
in Fig. \ref{Fig1}. 
The simplest case is found when the two components are rather 
similar like argon and krypton~\cite{schouten}. In a mixture of such atoms one find that the two 
critical points of the pure systems are connected by a line of critical points, 
well separated in density from the solidification 
transition: in this case, the two components are fully miscible in all the fluid phase.
This kind of transition will be referred to as ``liquid-vapour". 
In many, less simple, molecular fluids, like water/phenol mixtures~\cite{schouten}, 
a new phase transition may well arise in the high density  region, leading to the 
``demixing" between two liquid phases that differs in concentration. This transition 
ends in a new line of critical points and the phase diagram displays two lines of critical points, 
one in the small density region and one at large density. When the two components are 
more strongly dissimilar, like with Ne-Kr or He-Xe mixtures~\cite{schouten}, again one finds two 
lines of critical points but the topology of the phase diagram is different from the 
previous one because the two critical points of the pure components are no longer connected 
and two disjoint lines depart from them. One line moves toward large density transforming 
a liquid-vapor transition into a kind of demixing one while the other line terminates in an end point. 
The different character of the order parameter at any point along a critical line
can be fruitfully identified by an arrow in the 
density/concentration plane, whose direction provides the exact blending of density and concentration
fluctuations in the order parameter of the transition.  
Even more complex phase diagrams can be found, and up to 
six principal types of phase diagrams~\cite{rowlinson} have been classified for binary 
mixtures of small molecules. Not much attention has been given to which extent such 
richness of phase diagrams can be found in the area of colloidal suspensions or of other 
soft matter systems. This is the topic we address in the present paper. 

We consider a mixture of small non-adsorbing polymers and of spherical colloids 
in a solvent such that the colloids can be assimilated to hard spheres. 
Such polymers give rise to a short-range attraction between the colloids via the 
so-called depletion interaction~\cite{likos}. 
When colloids are monodisperse such an attraction gives rise to a phase transition 
between two phases, one rich and one poor in colloids and this transition mimics 
the liquid-vapor transition of a simple fluid, with the concentration 
of polymers playing the role of inverse temperature. When the range of the effective 
attraction is small enough compared to the colloid diameter, this 
transition can be metastable with respect to a direct transition between a low-density 
phase and a solid phase at high density of colloids~\cite{frenkel}. 
We consider now a similar system when colloids are not monodisperse but are a 
mixture of spheres of two diameters, $\sigma_1$ and $\sigma_2$. Since colloids are never monodisperse, 
our model is appropriate when the size distribution of colloids is bimodal with maxima 
around two sizes with a dispersion that is small compared to the difference $|\sigma_1-\sigma_2|$. 
In essence we have a three component system, colloids of two comparable sizes 
and small polymers in a background good solvent. When the polymer degrees 
of freedom are averaged out three depletion interactions arises, the two between 
colloids of the same size and that between colloids of different size. 
The depletion interaction between two colloids of different diameter 
is known~\cite{fantoni} in the framework of the Asakura-Oosawa (AO) model in which the 
inter-polymer interaction is neglected and the polymer-colloid interaction is assumed 
to be hard sphere. However, how such binary AO model affects the phase diagram of the 
system has not been studied and this is our main scope in this paper. In addition, 
recently, it has been pointed out that, by relaxing the hard 
core hypothesis, a strong enhancement in the strength of the depletion interactions can 
be achieved~\cite{rovigatti}.
We have then extended the colloidal mixture/polymer model when polymer-colloid interaction is 
represented by a repulsion over a finite range and it is not infinitely sharp 
as in the AO model. Based on such interaction models we ask if the mixture of 
two colloids of comparable size has only a liquid-vapor transition like in the 
monodisperse system or if also a demixing transition is present and which is the 
topology of the phase diagram. To this end we have studied the model on the basis 
of a suitable mean-field approximation and of the mapping onto the 
Adhesive Hard Sphere mixture model~\cite{baxter} 
determining the phase diagram as function of the size ratio $\sigma_2/\sigma_1$, of the size of 
the polymers $\xi$ and of the softness of the polymer-colloid interaction. 
We find that a demixing phase transition can indeed be present and all three topologies 
of phase diagram displayed in Fig. \ref{Fig1} can be realized depending on the model parameters. 

In Section II we first sketch the derivation of the two body effective interaction between 
two particle immersed in an ideal depletant for arbitrary colloid-depletant interaction, 
thereby generalizing the 
Asakura-Oosawa model. In Section III we motivate the use of a modified Mean Field 
free energy to investigate the phase diagram of the colloidal suspension. In Section IV
the results are shown examining the extension to binary fluids of the Asakura-Oosawa model first, 
and then studying the effects of softer colloid-polymer interactions. Some conclusion and perspective 
is offered in Section V.  

\section{The Soft Polymer Model: effective interactions} 
We provide a short derivation of the effective interaction between 
two colloidal particles, represented by hard-spheres of unequal diameters 
$\sigma_1$ and $\sigma_2$ respectively, immersed into a non-adsorbing polymer fluid. 
Following a well established procedure~\cite{likos}, polymers are 
regarded as non interacting point particles, which however feel a 
repulsive force due to the presence of the colloidal particles. 
Usually such an interaction is represented by the 
hard core constraint 
but, in view of future generalizations, here we allow 
for an arbitrary pair potential $v^{cd}_i(r)$ between a colloid of type $i=1,2$ 
and the depletant particles (polymers). The effective colloid-colloid 
interaction $v_{ij}^{eff}(r)$ induced by the presence of the depletant can be evaluated 
exactly in this model. The calculation is most easily performed 
in the semi-grand canonical ensemble~\cite{dijkstra} where the 
partition function of two colloids, of species $i$ and $j$, in a reservoir of depletant particles
is written as: 
\begin{equation}
\Xi_{ij} = \int d\bR_i \,d\bR_j \, e^{-\beta v^{cc}_{ij}(\bR_i-\bR_j)} \, \Xi_{ij}(\bR_i-\bR_j) 
\label{xi}
\end{equation} 
where $v^{cc}_{ij}(\bR)$ is the bare potential between the two colloids, $\beta=(k_BT)^{-1}$ and 
\begin{equation}
\Xi_{ij}(\bR_i-\bR_j) = \sum_N \frac{z^N}{N!} 
\int d\br_1\cdots d\br_N \,e^{-\beta \sum_n \left [v^{cd}_i(\br_n-\bR_i) + v^{cd}_j(\br_n-\bR_j)\right ]}
\label{xid}
\end{equation} 
is the depletant partition function, 
with $z = e^{\beta\mu}\Lambda^{-3}$ the fugacity of the ideal gas, $\mu$ its chemical potential 
and $\Lambda$ the De Broglie thermal wavelength. The partition function (\ref{xid}) is readily evaluated 
with the result 
\begin{equation}
\log \Xi_{ij}(\bR_i-\bR_j) = z \int d\br \, e^{-\beta \left [v^{cd}_i(\br-\bR_i) + v^{cd}_j(\br-\bR_j)\right ]}
\label{xid2}
\end{equation}
leading to an effective colloid-colloid interaction 
\begin{equation}
\beta v^{eff}_{ij}(\bR_i-\bR_j) = \beta v_{ij}^{cc}(\bR_i-\bR_j) -z
\int d\br \, \left \{e^{-\beta \left [v_i^{cd}(\br-\bR_i) + v_j^{cd}(\br-\bR_j)\right ]} -1\right \}
\label{vef}
\end{equation}
This result is exact irrespective of the specific form of the bare 
colloid-colloid ($v_{ij}^{cc}$) or colloid-depletant interaction $v_{ij}^{cd}$. 
Is it customary to write Eq. (\ref{vef}) in terms of the ``depletant reservoir density'' $\rho_d$
which in fact just coincides with the fugacity $z$~\cite{hansen}. By subtracting an irrelevant 
additive constant to the interaction, the effective inter-colloid 
potential can be written in the final form~\cite{rovigatti}: 
\begin{equation}
\beta v_{ij}^{eff}(\bR_i-\bR_j) = \beta v_{ij}^{cc}(\bR_i-\bR_j) -\rho_d
\int d\br \, \left [ e^{-\beta v^{cd}_i(\br-R_i)} -1\right ] 
\left [ e^{-\beta v_j^{cd}(\br-R_j)} -1\right ] 
\label{veff}
\end{equation}
The state dependent additive contribution, omitted in Eq. (\ref{veff}), does not affect the 
phase behaviour of the mixture, as explained in Sec III of Ref.\cite{dijkstra2}.
We remark that the form of the effective interaction derived here 
is correct when just two colloids are present in the system. 
In general, many body forces may appear at finite density of colloids\cite{dijkstra}, but, when 
the range of the colloid-depletant interaction is sufficiently small, these terms play a minor 
role in determining the phase diagram of the system and can be safely neglected. 
The expression for the effective interaction can be evaluated analytically
in the case of a binary AO mixture, where $v_i^{cd}(r)$ is a hard sphere interaction with range
$D_i=\frac{\sigma_i+\xi}{2}$ and $\xi$ is a parameter identifying the polymer size, as the polymer 
gyration radius:
\begin{equation}
\beta v_{ij}^{eff}(r) = \cases{ 
+\infty & for $r<\sigma_{ij}$ \cr
-\rho_d\, \nu_{ij}(r) & for $\sigma_{ij}<r<\sigma_{ij}+\xi$ \cr
0 & for $r>\sigma_{ij}+\xi$ \cr}
\end{equation}
where $\sigma_{ij}=\frac{\sigma_i+\sigma_j}{2}$ and
the function $\nu_{ij}(r)$ is given by \cite{fantoni}:
\begin{equation}
\nu_{ij}(r) = \frac{\pi}{12} \left [ r^3 -6\,(D_i^2+D_j^2)\,r + 8\,(D_i^3+D_j^3)
-3\,(D_i^2-D_j^2)\frac{1}{r}\right ] 
\end{equation}
In soft matter physics, when the effective potential between colloids is short ranged, 
a useful parameter measuring 
the strength of the interaction is given by the second virial coefficient or, equivalently,
by the so called Baxter's stickiness parameter
$\tau$~\cite{baxter} defined, in terms of the pair potential $v(r)$ by 
\begin{equation}
\int d\br\,\left [ e^{-\beta v(r)} -1\right ] 
= -\frac{4\pi}{3}\,\sigma^3\,\left [ 1-\frac{1}{4\tau}\right ]
\label{tau}
\end{equation}
where $\sigma$ is the hard core diameter. According to the empirical Noro-Frenkel law of
corresponding states, the liquid-vapour coexistence curve in a fluid interacting via a short 
range potential is remarkably independent of the details of the interaction if 
the temperature is measured in terms of $\tau$. 

\begin{figure}
\includegraphics[width=8cm]{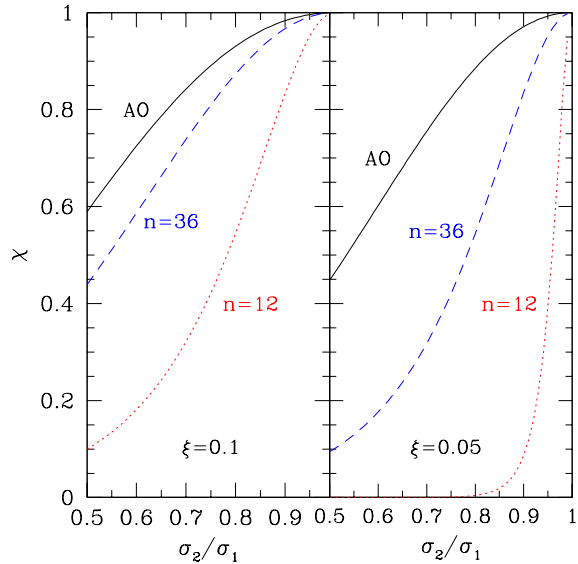}
\caption{Demixing ratio as a function of the colloid size ratio at a depletant volume fraction 
$\eta_d=\frac{\pi}{6}\rho_d\xi^3=0.1$. 
Left panel $\xi=0.1\,\sigma_1$. Right panel 
$\xi=0.05\,\sigma_1$. Full black curves: AO potential. 
Dashed blue curves: power law colloid-depletant interaction with 
softness exponent $n=36$. Dotted red curves: same with $n=12$. 
}
\label{Fig2}
\end{figure}

An estimate of the propensity to demixing of the AO mixture can be given in terms of the stickiness 
parameters $\tau_{ij}$ of the three effective potentials, as given by 
Eq. (\ref{tau}): If the spatially integrated strength of the unlike 
interaction is smaller then the average of the interactions between particles of 
the same species, we expect that, at low temperatures, a large portion of the phase diagram will 
display coexistence between phases of considerably different concentration. It is then natural to
define the demixing ratio in terms of the integrated strengths of the Boltzmann factors of
the depletion interactions:
\begin{equation}
\chi=\frac{2\,\sigma_{12}^3\tau^{-1}_{12}}{\sigma_1^3\tau^{-1}_{11}+\sigma_2^3\tau^{-1}_{22}}
\label{chi}
\end{equation}
In the following, the case $\chi <1$, i.e. when demixing is favored, will be named ``sub-additive".
A plot of this quantity as a function of the size ratio is shown in Fig. \ref{Fig2} for two 
choices polymer size $\xi$.
By lowering the size of the depletant, the tendency towards demixing is enhanced. 

\begin{figure}
\includegraphics[width=8cm]{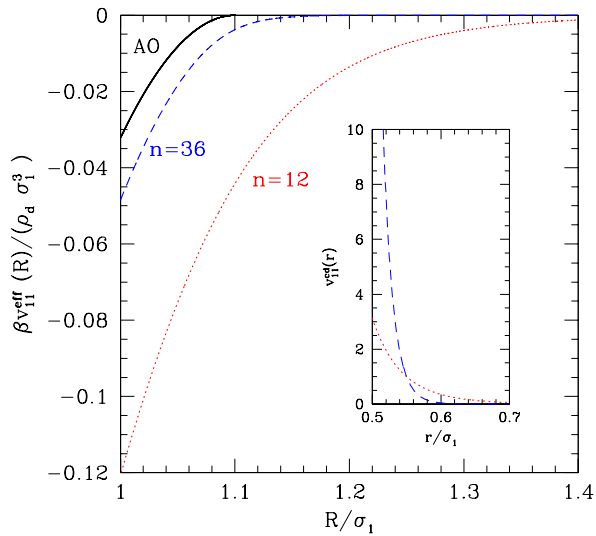}
\caption{Effective interaction between two hard spheres of diameter $\sigma_{1}$ immersed in
a fluid of non interacting depletant particles of density $\rho_d$. 
The colloid polymer interaction is defined by $\xi=0.1\,\sigma_{1}$
and by a softness exponent $n$. Full black curve
AO potential (i.e. $n\to\infty$); dashed blue curve $n=36$; dotted red curve $n=12$.
In the inset the bare colloid-depletant potential with $n=36$ (dashed blue) is compared 
with the choice $n=12$ (dotted red)
}
\label{Fig3}
\end{figure}

We may also consider more general colloid-depletant interactions by relaxing the hard sphere constraint
and introducing a suitable control parameter. As an example, we have investigated a 
{\it soft core} power-law potential parameterized by the 
characteristic length-scales $D_i$ for each colloid species $i=1,2$ and by the softness 
exponent $n$~\cite{rovigatti}:
\begin{equation}
\beta v^{cd}_i(r) = \left ( \frac{D_{i}}{r}\right )^n
\label{vcd}
\end{equation}
The colloid-colloid bare interaction is modeled as that of 
additive hard spheres, leading to the effective potentials shown 
in Fig. \ref{Fig3} for a few representative choices of the parameters.   

\begin{figure}
\includegraphics[width=6cm]{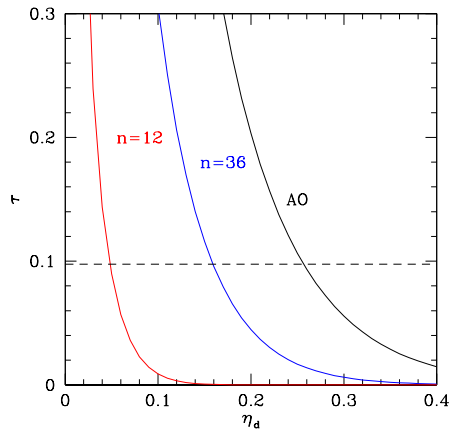}
\caption{Stickiness parameter $\tau$ as a function of the depletant volume fraction 
for three choices of the softness exponent: $n=\infty$ corresponds to the Asakura-Oosawa interaction, 
$n=36$ and $n=12$. The depletant size parameter is  $\xi=0.1\, \sigma_{1}$.
The dashed line sets the onset of the liquid-vapour transition and the crossing point sets the amount of 
depletant needed to drive phase transition in the colloidal suspension.
}
\label{Fig4}
\end{figure}

The presence of a power-law repulsive decay in the polymer-colloid interaction, even with a remarkably 
large value of the softness exponent $n$, gives rise to an attractive 
tail in $v^{eff}_{ij}(r)$, that may lead to significant changes in the 
phase diagram of the colloidal mixture. 
To quantify this effect, it is instructive to plot Baxter's 
stickiness parameter $\tau$ as a function of the physical depletant 
volume fraction $\eta_d = \frac{\pi}{6}\xi^3 \rho_d$
for a few choices of the softness exponent. According to the analytical 
solution of the PY equation~\cite{baxter}, a one component fluid is known to undergo a liquid-vapour transition when
$\tau$ reaches its critical value $\tau_c\sim 0.0976$. In Fig. \ref{Fig4} we can see that the 
amount of depletant needed to trigger the liquid-vapour transition is strongly reduced in going from 
the original AO model to a colloid-depletant potential of the form (\ref{vcd}) with $n$ as large as $n=36$.
A reduction in the polymer size parameter $\xi$ lowers the critical depletant volume fraction $\eta_d$ 
in a roughly linear way. 

On this basis we expect that, also in a colloidal mixture, the occurrence of colloid-depletant 
interactions of finite strength is going to significantly affect the topology of the 
phase diagram, which is 
known to be very sensitive to the relative strengths of the interactions, 
as shown in Fig. \ref{Fig2}. By increasing the softness of the colloid-depletant 
interaction, the tendency towars demixing is enhanced.  

\section{Free energy models for a colloidal mixture}
\label{mf}
The theoretical investigation of the phase diagram of binary fluids must rely on 
some kind of approximation. For effective interactions whose range is short
with respect to the colloid diameter, standard mean field 
approaches are affected by serious inaccuracies. Even in the
simpler case of one component fluids, van der Waals-like theories imply the law of 
corresponding states 
and lead to quantitative discrepancies with respect to 
the results of numerical simulations~\cite{frenkel}. The situation is considerably worse when 
the standard mean field theory is extended to binary fluids, because the different 
topologies of the phase diagram which a mixture may display depend on a delicate balance
between entropic and energetic contributions to the free energy. 
Unfortunately, such a balance is not captured by any free energy linear in the interactions.
In order to circumvent this
problem, a simple generalization of the standard mean field approach has been rediscovered several times
in liquid state theory~\cite{pini}: it simply amounts to assume that, outside the core, the
direct correlation function coincides with the Mayer function,
rather than with the potential divided by $-k_BT$. This simple way to modify the usual Random 
Phase Approximation~\cite{hansen} (RPA) at least displays the correct low density limit and
satisfies the Noro-Frenkel criterion~\cite{noro} providing 
a clear improvement on standard RPA with a modest computational effort. 
By comparing the structure factor predicted by this modified RPA with computer simulations 
or with the results of integral equations, a good agreement is found at low and moderate 
densities, while in the high density/low temperature region the modified RPA tends to 
overestimate the peack height in $S(k)$. 

A different point of view, for dealing with short range potentials, is to resort to the 
celebrated integral equations of liquid state theory, like HNC, PY or MSA~\cite{hansen}. 
Few of them also allow for an analytic solution in the ``adhesive sphere" 
limit, i.e. in the limit of 
vanishing range of the attractive potential~\cite{baxter,giacometti}. Unfortunately, the
analytic solution shows that PY does not lead to a full spinodal curve: this integral equation
has no solution on one side of the critical point, leading to difficulties in extracting the 
thermodynamics from the knowledge of the correlation function. On the other hand, when 
applied to mixtures, MSA for sticky potentials is known to be affected by a serious inconsistency
which does not allow to obtain a free energy function via the compressibility route~\cite{giacometti2}.
Finally, a differential approach to the phase equilibria in mixtures (HRT), 
which proved successful for molecular fluids, has not yet been implemented 
for short range potentials~\cite{aplr}.  

The phase diagram of our colloidal mixture, interacting via the effective pairwise potential $v^{eff}(R)$,
has been investigated starting from a mean-field like free energy density of the form:
\begin{equation}
A(\rho_1,\rho_2,\rho_d) = A_{HS}(\rho_1,\rho_2) -\frac{k_BT}{2}\,
\sum_{i,j=1}^{2}\,\rho_i\rho_j \int d\br \left [ e^{-\beta v_{ij}^{eff}(r)} -
e^{-\beta v_{ij}^{cc}(r)} \right] 
\label{gmf}
\end{equation}
where $\rho_1$ and $\rho_2$ are the number densities of the two species and
$A_{HS}(\rho_1,\rho_2)$ is the free energy density of a hard sphere mixture,
well represented by the Mansoori-Carnahan-Starling expression~\cite{mansoori}.
Here we assumed a bare inter-colloid potential $v^{cc}_{ij}(r)$ of the hard core form.
This Ansatz, extends in a straightforward way the previously mentioned modified RPA
for one component fluids~\cite{pini}.
It can be seen as a generalization of the standard Mean Field 
expression for the Helmholtz free energy density: 
\begin{equation}
A_{MF}(\rho_1,\rho_2,\rho_d) = A_{HS}(\rho_1,\rho_2) +\frac{1}{2}\,
\sum_{i,j=1}^{2}\,\rho_i\rho_j \int d\br \left [v_{ij}^{eff}(r)-v^{cc}_{ij}(r)\right ] 
\label{mf0}
\end{equation}
where the entropic contribution coincides with that of a pure hard sphere 
binary fluid, and equals $A_{HS}$, while the internal energy is evaluated in 
Hartree approximation~\cite{hansen}. However, this Mean Field approximation, being linear in 
the attractive part of the interaction ($v_{ij}^{eff}$), can be justified only in the limit 
of weak and long ranged potential (the so called Kac limit), which is not the case of interest
(see  Fig. \ref{Fig4}). To improve this approximation, maintaining the physical content of 
Mean Field Theory, we can look at the low density limit, where the free energy is known to 
behave as 
\begin{equation}
A(\rho_1,\rho_2,\rho_d) \to A^{id}(\rho_1,\rho_2) -\frac{k_BT}{2}\,
\sum_{i,j=1}^{2}\,\rho_i\rho_j \int d\br \left [ e^{-\beta v_{ij}^{eff}(r)} - 1 \right] 
\end{equation}
The easiest way to enforce this asymptotic behaviour into an approximation of the form (\ref{mf0})
is given by Eq. (\ref{gmf}), which is in fact our strategy. 
An intriguing direct consequence of such an approximation is
that the Noro-Frenkel law of corresponding states \cite{noro} 
is here extended to the case of binary mixtures.
In fact, within this Mean Field theory,  the phase diagram is independent of
the details of the interactions, provided the temperature is expressed in terms of Baxter's $\tau_{ij}$ 
parameters (\ref{tau}). 
Note that when, according to our definition (\ref{chi}), $\chi=1$ the attractive interaction just 
couples the total density $\rho_1+\rho_2$ in Eq. (\ref{gmf}) and demixing will not be favoured. 

In the same spirit of Eq. (\ref{gmf}), we can evaluate pair correlations in the mixture by
use of a modified RPA, expressing the direct correlation functions of the mixture as:
\begin{equation}
c_{ij}(r) = c_{ij}^{HS}(r) + e^{-\beta v_{ij}^{eff}(r)} - 1
\label{cr}
\end{equation}

Starting from the free energy (\ref{gmf}), 
the critical lines are determined by the vanishing of the Hessian determinant of $A(\rho_1,\rho_2)$:
\begin{equation}
\det \,\frac{\partial^2 A(\rho_1,\rho_2,\rho_d)}{\partial \rho_i\partial\rho_j} = 0 
\end{equation}
implying that one eigenvalue of the Hessian matrix vanishes. Let us call 
$\zeta_1$ and $\zeta_2$ 
the two linear combinations of $\rho_1$ and $\rho_2$ which diagonalize the Hessian matrix
at a given point of the spinodal line and let us denote by $\lambda_1$ the vanishing eigenvalue.
The critical point is identified by the condition 
$\frac{\partial^3 A(\zeta_1,\zeta_2,\rho_d)}{\partial \zeta_1^3}=0$,
while the convexity requirement of the free energy at the critical point
gives rise to two additional conditions:
\begin{eqnarray}
\frac{\partial^2 A(\zeta_1,\zeta_2,\rho_d)}{\partial \zeta_2^2} &>& 0  \\
\left [ \frac{\partial^3 A(\zeta_1,\zeta_2,\rho_d)}{\partial \zeta_1^2\partial \zeta_2} \right ]^2 
-\frac{1}{3} \frac{\partial^2 A(\zeta_1,\zeta_2,\rho_d)}{\partial \zeta_2^2} \,
\frac{\partial^4 A(\zeta_1,\zeta_2,\rho_d)}{\partial \zeta_1^4} &<& 0
\end{eqnarray}
These inequalities allow to detect spurious singularities selecting only the
stable transition lines. The two ``proper" variables $\zeta_1$ and $\zeta_2$ 
respectively identify the strong and weak fluctuation directions.
It is instructive to represent the strong fluctuation direction by an arrow in
the $(\rho_1,\rho_2)$ plane~\cite{mix}.

\section{Phase diagrams of the colloidal mixture}
We numerically investigated the solution of the mean field equations 
discussed in Section \ref{mf} by varying the model parameters.
The critical lines and the strong fluctuation direction
are shown in the volume fraction vs. concentration plane
($\eta=\frac{\pi}{6}(\rho_1\sigma_1^3 + \rho_2\sigma_2^3)$; $x=\eta_1/\eta$).

\begin{figure}
\includegraphics[height=7cm,width=7cm,angle=0]{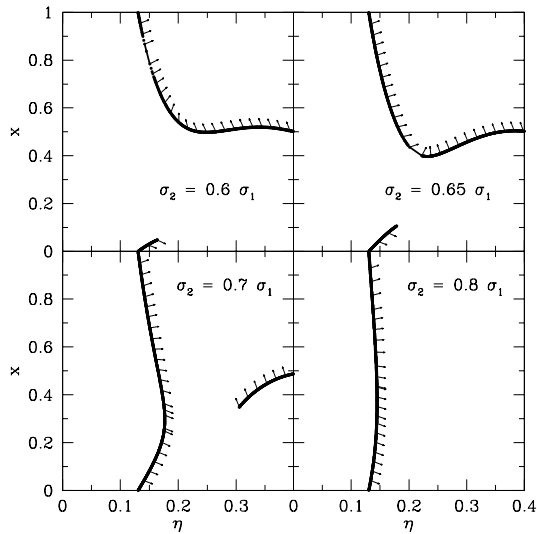}
\caption{Projection of the critical lines in the volume fraction/concentration plane for a 
hard sphere colloidal mixture with an AO depletant of diameter $\xi=0.1\sigma_1$. The arrows 
show the character of the order parameter: 
an horizontal arrow denotes a density fluctuation and a vertical arrow a concentration fluctuation. 
}
\label{Fig5}
\end{figure}

We first consider a binary fluid of hard particles of diameters $\sigma_1$ and $\sigma_2$ 
with standard AO effective interaction. As previously noted, in the pure system (i.e. at concentration
$x=0$ or $x=1$) we expect a liquid-vapour transition at colloid volume fraction $\eta\sim 0.12$ and 
depletant packing fraction $\eta_d$ larger than $\eta_d \sim 0.26$. The projections of the critical 
lines are shown in Fig. \ref{Fig5} for $\xi=0.1\,\sigma_1$ and few choices of the colloid size ratio.
For a size ratio lower than $0.65$ the transition changes character along the 
critical line, from liquid-vapour at high concentration
of large particles to demixing at equimolar concentrations: The phase diagram falls
in the class III previously discussed. By increasing $\sigma_2/\sigma_1$, the topology of the critical 
lines is modified, moving through type II for size ratio $0.7$ and finally to type I, 
i.e. pure liquid-vapour, 
above a $\sigma_2\sim 0.8\,\sigma_1$. Note that the transition between type II and type I is not sharp and the 
demixing line appears to grow from the high density region, emerging from the solid phase (not shown 
in these diagrams).  

The possibility to obtain an efficient fractionalization of colloids of similar   
sizes by adding polymers to the suspension 
is however hampered by the large depletant volume fraction required to trigger
phase separation ($\eta_d\,\sim 0.3$). 
We may also add that, at those densities, considering polymers as an ideal 
gas is likely to be rather inaccurate. 

\begin{figure}
\includegraphics[height=7cm,width=7cm,angle=0]{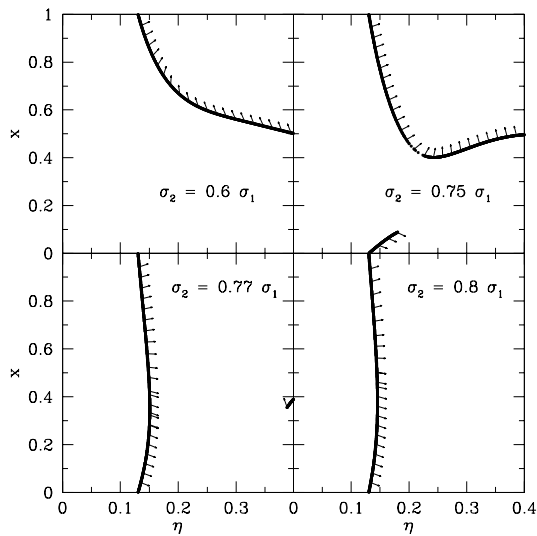}
\caption{Same as Fig. \ref{Fig5} for $\xi=0.05\,\sigma_1$. 
}
\label{Fig6}
\end{figure}

\begin{figure}
\includegraphics[height=5cm,width=10cm,angle=0]{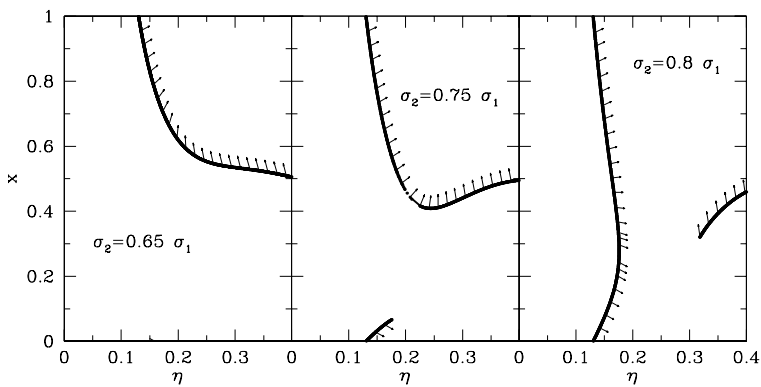}
\caption{Same as Fig. \ref{Fig5} for a colloid-depletant interaction with 
softness exponent $n=36$ and $\xi=0.1\, \sigma_1$. 
}
\label{Fig7}
\end{figure}

By reducing the polymer gyration ratio to $\xi=0.05\,\sigma_1$, the depletion mechanism is more 
efficient and the demixing transition between the two topologies of the phase diagram 
occurs at larger 
colloid size ratio, as can be seen in Fig. \ref{Fig6} . Also the amount of depletant required
to trigger the instability is lowered down to $\eta_d\sim 0.2$, which still corresponds to a 
very dense suspension ($\eta+\eta_d \sim 0.4-0.5$).
Reducing the hardness of the colloid-polymer interaction leads to considerable effects on the 
phase diagram: the sequence of topologies 
is shifted towards higher size ratios, showing that softness enhances the propensity 
towards demixing, and the required polymer density is reduced, 
as expected on the basis of Fig. \ref{Fig4}.  As an example, we 
show in Fig. \ref{Fig7} the sequence of phase diagrams which are obtained for $n=36$ and $\xi=0.1 \sigma_1$.
The critical depletant volume fraction along the transition lines 
varies smoothly around $\eta_d\sim 0.2$. 
Also in this case, by reducing the polymer size $\xi$, the depletant density further lowers 
and the demixing transition occurs at a size ratio as high as $0.85$, as shown in Fig. \ref{Fig8}. 
\begin{figure}
\includegraphics[height=5cm,width=10cm,angle=0]{fig8.eps}
\caption{Same as Fig. \ref{Fig5} for a colloid-depletant interaction with 
softness exponent $n=36$ and $\xi=0.05\,\sigma_1$. The red arrows in the left and right panel 
show the critical points marked by a red dot in Fig. \ref{Fig9}
}
\label{Fig8}
\end{figure}
Here, the amount of depletant required to induce phase separation drops to $\eta_d \sim 8\%$,
showing that adding a small amount of polymer to a binary mixture of hard spheres may give rise to
sharp demixing provided the polymer-colloid interaction is not too harsh. 
By further softening the colloid-polymer interaction the effect becomes more and more pronounced.
A softness exponent $n=12$ gives the same sequence of phase diagrams but the amount of depletant required
to drive the demixing of the colloids is lower than $2\%$. 

\begin{figure}
\includegraphics[height=6cm,width=6cm,angle=0]{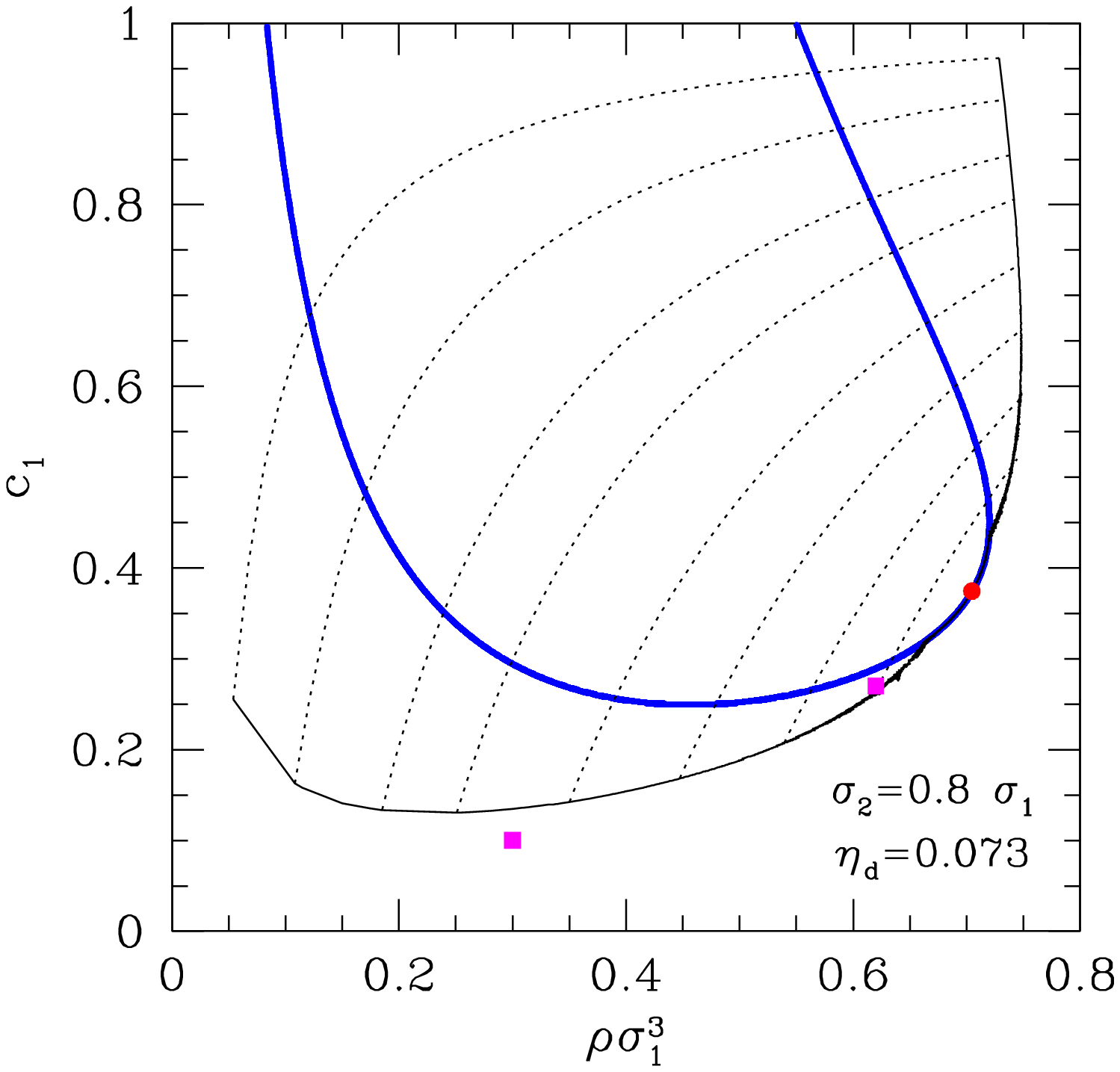}
\includegraphics[height=6cm,width=6cm,angle=0]{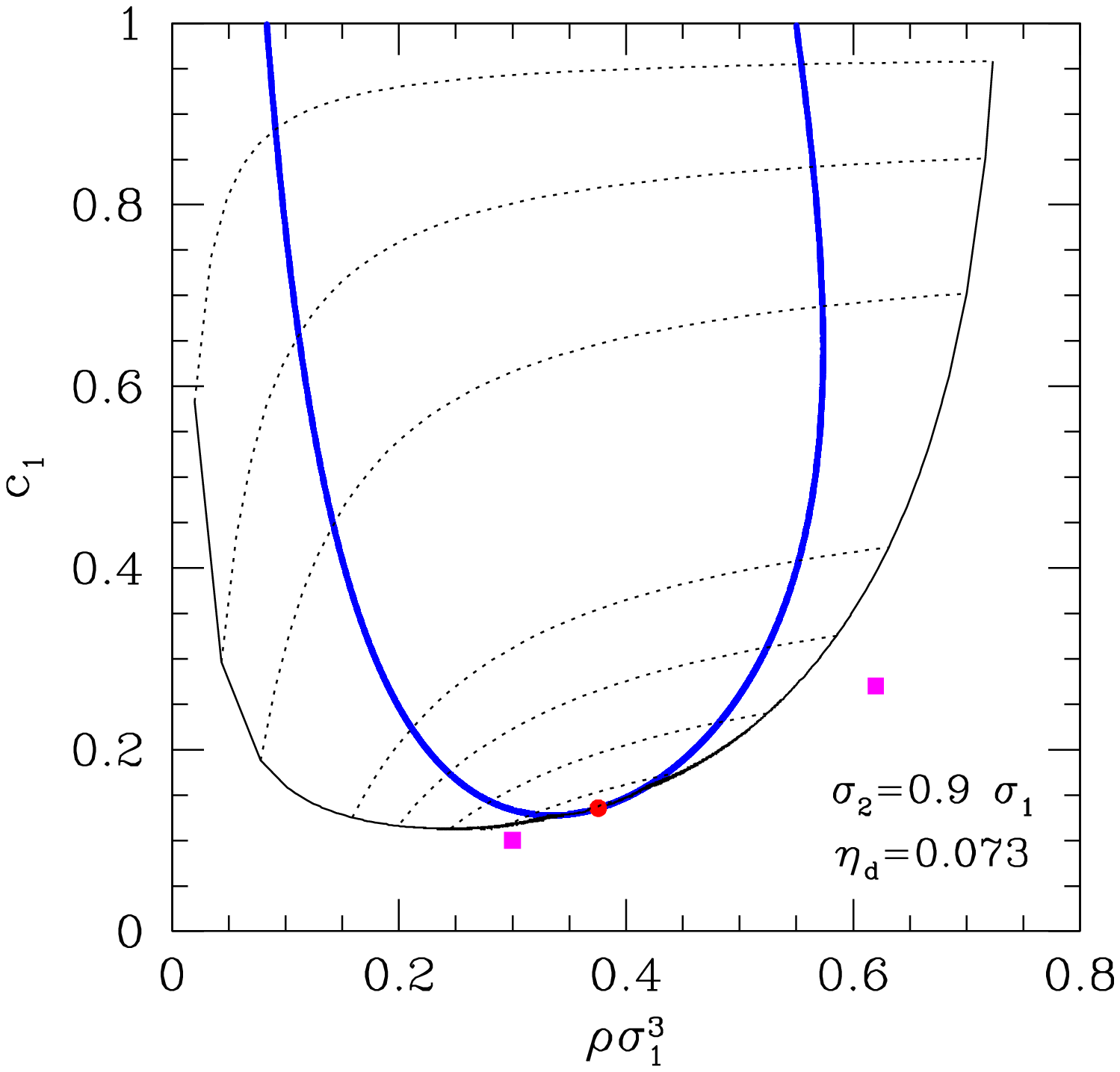}
\caption{Two slices to the phase diagram in the $(\rho,c)$ plane 
for two choices of the colloid size ratio. 
Here $\xi=0.05\sigma_1$ and $n=36$. The blue line is the spinodal, the 
full black line is a portion of the coexistence curve and the red dot is 
the critical point. The dotted lines are 
tie lines, joining the coexisting densities and compositions.  
The squares (magenta) show the state points where correlations have been depicted in Figs. 
\ref{Fig10},\ref{Fig11}.
}
\label{Fig9}
\end{figure}

The nature of the transition can be better 
appreciated by contrasting the shape of the transition lines in the total 
density $\rho=\rho_1+\rho_2$ vs.
concentration plane ($c_i=\frac{\rho_i}{\rho}$), at a given polymer volume fraction, 
for two different topologies of the 
phase diagram. As an illustrative example, in Fig. \ref{Fig9} we compare 
the case $\sigma_2=0.8\,\sigma_1$ 
with the case $\sigma_2=0.9\,\sigma_1$ for the choice $\xi=0.05\,\sigma_1$ and $n=36$ 
at $\eta_d=7.3\%$. 

As shown in Fig. \ref{Fig8}, although the change in size ratio 
is rather small ($10\%$), it is enough to switch between the two classes of 
phase diagrams: from type III to type II. 
In both cases the two-phase region fills a large portion of the density-concentration plane, 
but the nature of the transition changes. To visualize this difference it is instructive to 
follow the tie lines (dotted lines in the figures) which join  
the two coexisting phases at a given average density $\rho$ and concentration $c$. 
It is apparent how the topology of the phase diagram affects the composition of the coexisting phases:
In type III mixtures (like the case shown in the left panel), when the average total 
volume fraction is suitably chosen, the coexisting phases have both
large density {\it and} concentration mismatch, while in type I or II systems (right panel) 
the difference in concentration between the coexisting phases is moderate. 

Finally, we briefly discuss how the topology of the phase diagram reflects in particle correlations. 
The partial structure factors are a direct measure of density fluctuations
and can be expressed as the equilibrium average of a product of density operators of species $i$ and $j$:
$S_{ij}(\bk) = \frac{1}{N}\,\langle \hat\rho_i(\bk)\hat\rho_j(-\bk)\rangle$ and are 
related to the direct correlation functions by the Ornstein-Zernike relation~\cite{hansen}.
The structure factors are evaluated by use of the modified RPA approximation Eq. (\ref{cr}). 
To discriminate between demixing and liquid-vapour transitions in binary mixtures
it is convenient to introduce suitable linear combinations of $S_{ij}(\bk)$,
the Bhatia-Thornton structure factors \cite{bhatia},
which properly characterize the density and concentration fluctuations:
\begin{eqnarray}
\label{srr}
S_{\rho\rho} &=& 
c_1\,S_{11}(\bk)+c_2\,S_{22}(\bk)+2\,(c_1c_2)^{1/2}\,S_{12}(\bk) \nonumber \\
\label{scc}
S_{cc} &=&  
c_2^2\,c_1\,S_{11}(\bk)+c_1^2\,c_2\,S_{22}(\bk)-2\,(c_1c_2)^{3/2}\,S_{12}(\bk) \nonumber \\
\label{src}
S_{\rho c} &=& S_{c \rho} =
c_1\,c_2\,\left [S_{11}(\bk)-S_{22}(\bk)\right ]+(c_2-c_1)\,(c_1c_2)^{1/2}\,S_{12}(\bk) 
\end{eqnarray}
where $c_i=\rho_i/\rho$, $\rho=\rho_1+\rho_2$ are the number concentrations of the
two species and the total density, respectively.

\begin{figure}
\includegraphics[height=6cm,width=6cm,angle=0]{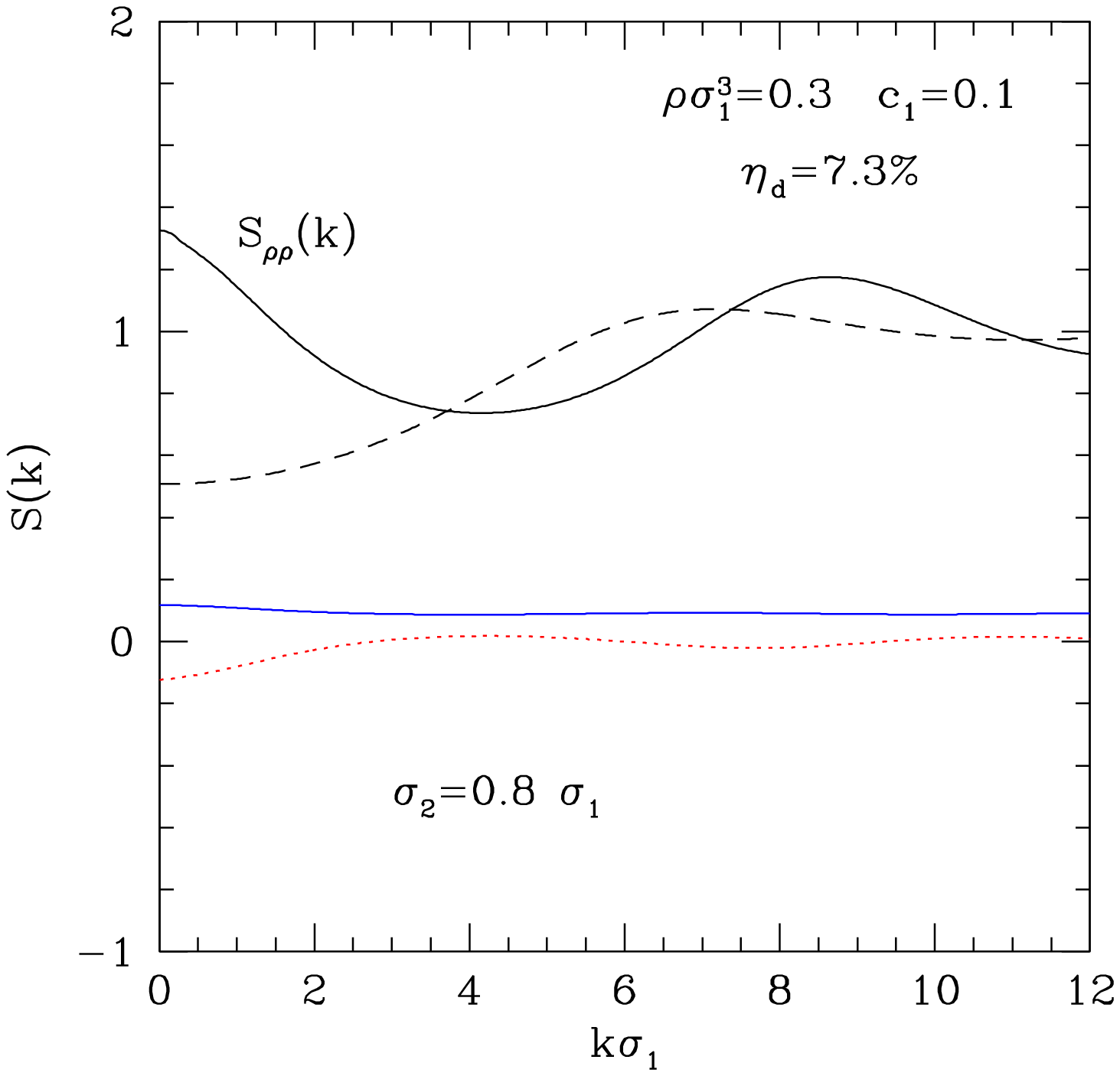}
\includegraphics[height=6cm,width=6cm,angle=0]{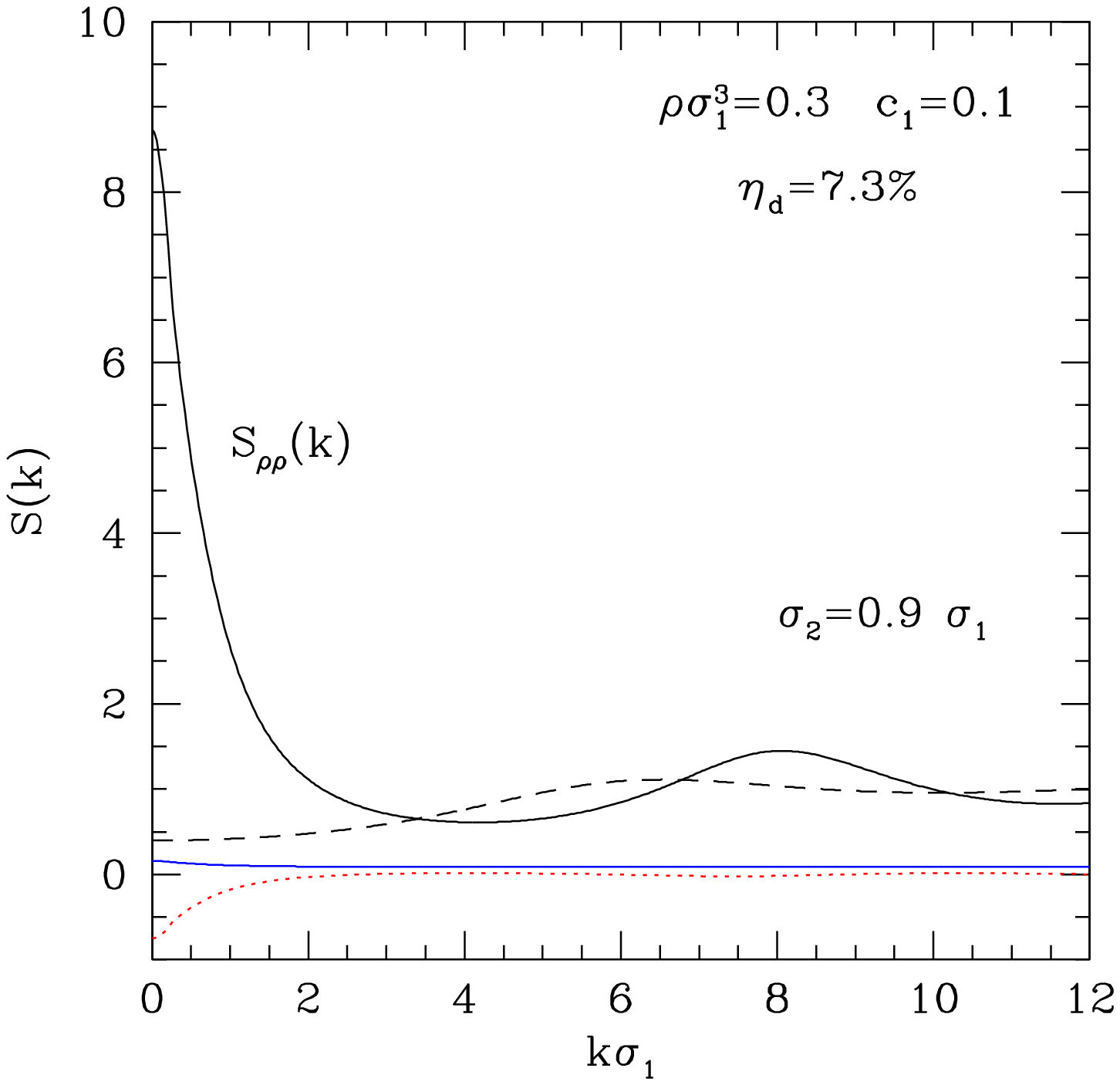}
\caption{Bhatia-Thornton partial structure factors for two choices of the colloid size ratio.  
The colloid-depletant repulsion is chosen to be power-law (\ref{vcd}): $n=36$, $\xi=0.05\sigma_1$. 
Density fluctuations (\ref{srr}) are shown by the full black curve. Concentration fluctuations (\ref{scc}) 
by a blue full line, while the dotted red curve gives the off-diagonal term. The density-density
structure factor for a hard sphere mixture at the same density and concentration is shown by
the dashed line. 
}
\label{Fig10}
\end{figure}

\begin{figure}
\includegraphics[height=6cm,width=6cm,angle=0]{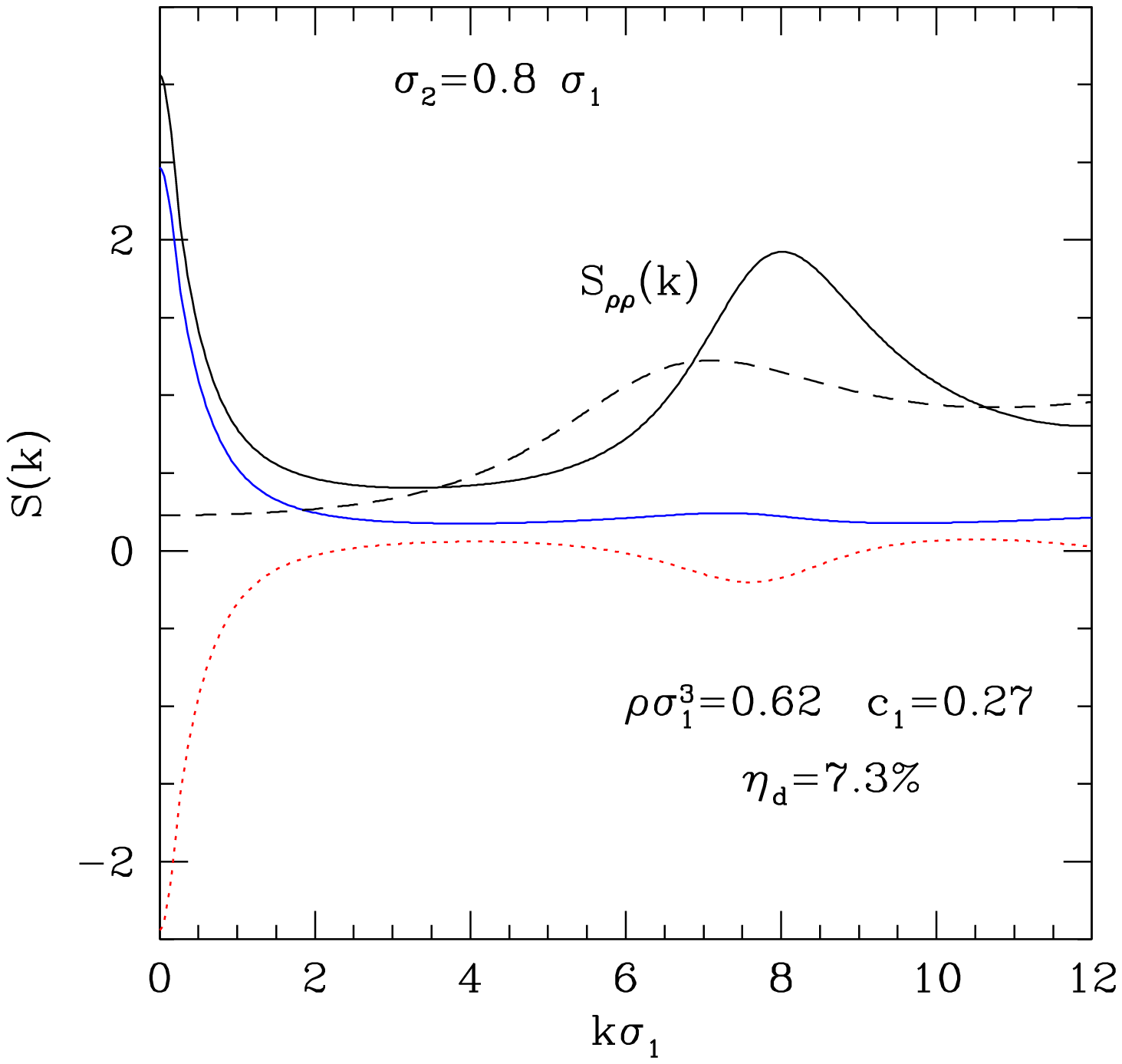}
\includegraphics[height=6cm,width=6cm,angle=0]{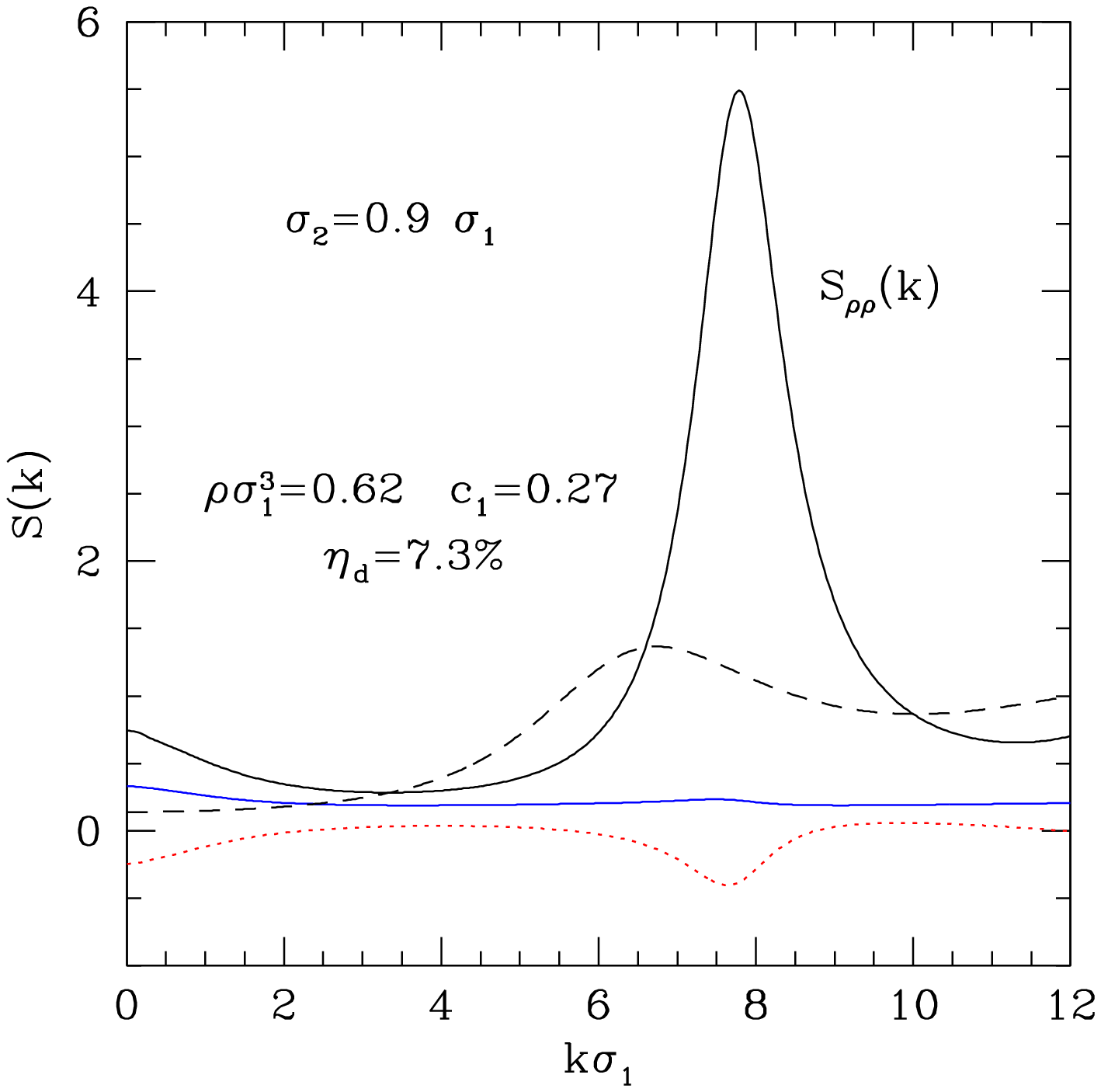}
\caption{Bhatia-Thornton partial structure factors for two choices of the colloid size ratio.  
Notation as in Fig. \ref{Fig10} 
}
\label{Fig11}
\end{figure}

As an illustrative example, the Bhatia-Thornton partial structure factors are shown at 
the same point in the phase diagram for two choices of the size ratio in Figs. \ref{Fig10}.
The chosen state point ($\rho\sigma_1^3=0.3$,$c_1=0.1$) 
is marked by a square in the sections of the density-concentration plane in Figs. \ref{Fig9}.
In both cases density correlations appear to be significant, even at a rather low volume fraction. 
Oscillations in $S_{\rho\rho}(\bk)$ reveal the presence of short range correlations among colloids.
At size ratio $0.9$, the presence of the liquid vapour critical point (shown in Fig. \ref{Fig9}) clearly affects density 
fluctuations in a large portion of the phase diagram, leading to a strong peak at vanishing 
wave vector. Concentration fluctuations are instead very weak and remarkably flat at this
point in the phase diagram. Figs. \ref{Fig10} also include the density-density structure 
factor (\ref{srr}) 
for a binary hard sphere mixture at the same density and concentration 
(concentration fluctuations are small and 
structureless and are not reported). Attractive interactions give rise to an enhancement of fluctuations 
at small wavevectors but also to a shift of the main peak towards higher wave vectors, reflecting 
the tendency to aggregation. 
These data can be contrasted with the results shown in Fig. \ref{Fig11}, referring to a state point at 
higher density ($\rho\sigma_1^3=0.6$,$c_1=0.27$), also marked by a square in Figs. \ref{Fig9}. 
In both cases density correlations acquire a significant structure, showing strong short range correlations
induced by depletion, 
particularly at  size ratio $0.9$ where a very high peak, probably overemphasized 
by our modified RPA approximation, develops. 
For $\sigma_2=0.8\,\sigma_1$, the closeness to the demixing
critical point enhances both density and concentration fluctuations at long wavelengths.
This finding agrees with previous remarks about the character of the demixing transition 
in these systems:
the coexisting phases in fact differ both in density and in concentration, implying a composite order 
parameter involving both kinds of fluctuations.   
In contrast, in a pure hard sphere mixture both density and concentration fluctuations are small.

\section{Conclusions}

The attractive nature of the so-called depletion interaction is well known as 
well as its effect in the densification of colloidal particles. 
Less well known is that depletion interaction has a demixing effect when 
colloids of unequal size are present.  We have shown how rich is the phase 
diagram of a mixture of colloids of different but comparable size in presence of 
a depleting agent like small non-adsorbing polymers. In fact the depletion interaction 
between two spheres of different size is sub-additive 
(see Eq. (\ref{chi}))
compared to the two interactions 
between pair of spheres of the same diameter. Therefore in a binary mixture of colloids two 
competing effects are present. On one hand the attractive nature of the depletion interaction 
brings in the tendency to phase separation of the colloids in two phases, one rich of colloids 
and one poor of colloids. On the other hand the sub-additivity of the depletion interaction 
between spheres of different diameter promotes phase separation of spheres of different diameter. 
All three main topologies of the phase diagram of a binary mixture described in the Introduction 
are found, depending on the values of the interaction parameters. The tendency to demixing 
is enhanced by an increasing asymmetry of the colloids and by a decreasing size of the 
polymers. In addition, a strong enhancement of demixing is found if the polymer-colloid 
interaction has some degree of softness and it is not the hard core one assumed in the 
AO model. In fact, we find that under suitable conditions even a concentration of polymers 
as low as few percent in volume fraction can be enough to cause demixing of the colloids.

We have considered mixtures of colloids of two different sizes. It is clear that our 
study is relevant also for real colloids when the size distribution $P(\sigma)$ is strongly bimodal, 
i.e. $P(\sigma)$ as function of the diameter $\sigma$ has two sharp maxima.  
We are not aware of experimental 
studies of similar colloidal mixtures in the fluid phase in presence of small polymers but only 
at very large volume fraction such colloids has been investigated \cite{palberg}
with the purpose of studying the crystalline phases. In this work a remarkable effect of the polymers 
on the kinetic of the crystallization was found. A binary mixture of hard spheres of similar 
diameters $\sigma_1$ and $\sigma_2$
is fully miscible in the fluid phase but upon solidification phase separation 
between different crystalline phases can be present \cite{frenkel}. 
Experiments \cite{bartlett} 
with mixtures of colloids of two sizes at high packing 
fraction indeed show solidification but usually a glass state is formed and the kinetic of 
crystallization is very slow. By adding small polymers in the system one 
finds \cite{palberg} 
a dramatic acceleration of the crystallization process 
and it has been conjectured that this is due to enhanced concentration fluctuations induced 
by the polymer. In that work the size ratio $\sigma_2/\sigma_1$ is $0.79$ and the polymer 
diameter ratio $\xi/\sigma_1=0.068$, 
values similar to those of our study. We can give now a rational basis to the finding of this 
experiment. Due to the sub-additivity of the depletion interaction between spheres of different 
diameter already in the fluid phase there is a tendency to demixing with enhanced concentration 
fluctuations so that the crystallization in solids with different composition should be facilitated. 
Actually a demixing transition might already be present in the fluid phase. An experimental 
verification of our prediction of demixing in the fluid phase and on the different 
topology of the phase diagram, depending on the interaction parameters, should be very interesting. 

Another interesting issue is how the demixing tendency of the depletion interaction acts at 
high packing fraction of binary colloids where gelation can set in the monodisperse system.

Our study is based on a rather simple free energy functional of the mixture, it is of a mean 
field form suitably generalized to treat strong and short range attractive forces that are 
typical of depletion by small particles. Our free energy is such that the Noro-Frenkel law of 
corresponding states is extended to binary mixtures, i.e. thermodynamic properties do not depend 
on the details of the three interaction potentials but only on the three stickiness parameters 
$\tau_{ij}$ ($i,j=1,2$) as given by Eq. (\ref{tau}). 
The theory could be improved but we believe that our basic findings are at least qualitatively correct. 
It should be interesting to check by simulation their accuracy on a quantitative footing. 

Often in the paper we have mentioned non-adsorbing polymers as depleting agent. 
It is clear that the same effects are expected when depletion is induced by other 
kind of small particles like small micelles. It remains to be studied how the sub-additivity 
of the depletion interaction is modified when the interaction between the depleting particles 
cannot be neglected as it is done in the AO model. An interesting result of our study is the 
prediction that the demixing effect of the depletion interaction is strongly enhanced when 
some softness is present in the polymer-colloid interaction. An experimental study of this 
aspect should be possible, for instance by coating the colloids by adsorbing polymers of different 
length or by temperature sensitive polymers so that the softness could be controlled by temperature.
Softening the direct colloid-colloid repulsive interaction may result either in favoring or in hampering 
particle demixing, depending whether the repulsive tail in $v^{cc}(r)$ prevails over the 
attraction induced by depletion. A realistic representation of the pair potential is therefore
required to investigate the competition between these effects.

It is a pleasure to dedicate this work to Jean-Pierre, whose deep insight and 
outstanding achievements permeate modern Liquid State Theory. 
We acknowledge fruitful discussions with P. Tartaglia, E. Zaccarelli, 
D. Ashton, N. Wilding and A. Giacometti.
One of us (L.R.) wants to thank Dipartimento di Fisica, Universit\'a
degli Studi di Milano, for some support to his research activity.

\end{document}